\documentclass[a4paper]{iopart}

\usepackage{iopams}
\usepackage[square,sort&compress,numbers]{natbib}
\usepackage{graphicx}

\eqnobysec

\begin{document}

\title{Classicalization by phase space measurements}

\author{Marduk Bola{\~n}os}

\address{Fakult{\"a}t f{\"u}r Physik, University of Duisburg-Essen,
         Lotharstr. 1, 47048 Duisburg, Germany}


\begin{abstract}
  This article provides an accessible illustration of the measurement
  approach to the study of the quantum-classical transition suitable
  for beginning graduate students. As an example, we apply it to a
  quantum system with a general quadratic Hamiltonian and obtain the
  exact solution of the dynamics for an arbitrary measurement
  strength.
\end{abstract}



\submitto{\EJP}


\section{Introduction}
\label{sec:introduction}

The study of the quantum-classical transition is an active field of
research that has provided important insights into the foundations of
quantum theory and plays a prominent role in the development of
quantum technologies \cite{Schlosshauer}.

Some of its achievements have been obtained through the measurement
approach, which consists in using the framework of generalized
measurements \cite{Busch_et_al} to model the dynamics of an open
quantum system, i.e. a quantum system in interaction with a large
number of quantum degrees of freedom, which are collectively called
environment. This has been used, for example, to develop a
quantitative assessment of the macroscopic character of a
superposition state based on the experimental observation of quantum
effects \cite{Nimmrichter}.

The concepts and methods used in this field are usually not familiar
to beginning graduate students. However, some of them have been
discussed in a pedagogical way in recent years
\cite{Case,LovettNazir,Pearle,XuLiMaLi}. The main contribution of this
article is to present an accessible illustration of the measurement
approach applied to an analytically tractable model for the
classicalization of a quantum system described by a general quadratic
Hamiltonian. This should be useful for instructors with an interest in
introducing graduate students to current research topics in quantum
mechanics.

The manuscript is organized as follows. In section 2, the
Heisenberg--Dirac formulation of quantum mechanics is briefly
discussed, with a focus on physical dimensions and algebraic
considerations. These two aspects will play an important role
throughout the article. We remark that the use of algebraic methods in
quantum mechanics has lead to a deeper understanding of nature, while
also offering an elegant and powerful framework to study a wide
variety of systems \cite{Woit,ThyssenCeulemans}. Section 3 contains a
brief account of the phase space formulation of quantum mechanics,
which allows one to establish a connection with classical
mechanics. In section 4 generalized measurements are defined and it is
shown how to construct a dynamical equation for the state of a system
subject to such measurement. Section 5 contains an example of
measurement-induced classicalization. In particular, we consider an
imprecise, simultaneous measurement of two canonically conjugate
observables of a quantum harmonic oscillator, which leads to the decay
of all the coherences in a superposition state.

\section{Heisenberg--Dirac quantum mechanics}
\label{sec:Heisenberg_Dirac_QM}

In analogy to classical Hamiltonian mechanics, the observable
quantities of an elementary quantum system are described by
operator-valued functions of two self-adjoint operators
$\mathsf{Q} = \mathsf{Q}^\dagger, \mathsf{P} = \mathsf{P}^\dagger$
satisfying the canonical commutation relation
\begin{equation}
  \label{eq:CCR_x_p}
  [\mathsf{Q}, \mathsf{P}] = i\hbar \mathsf{I}.
\end{equation}
The operators $\mathsf{Q}$, $\mathsf{P}$, and $\mathsf{I}$ (identity)
are the canonical basis of the Heisenberg Lie algebra
$\mathfrak{h}(3)$ \cite{KlimovChumakov}. It is customary to choose
units in which $\hbar = 1$, but this makes it difficult to verify that
an equation has the correct dimensions. For economy of notation it is
preferable to use the dimensionless basis of $\mathfrak{h}(3)$ given
by \cite{Serafini}
\begin{equation}
  \label{eq:CCR_ladder_basis}
\mathsf{a} = \frac{1}{\sqrt{2}}\left(\frac{\mathsf{Q}}{\lambda} + \frac{i}{\hbar}\mathsf{P}\lambda\right),\quad \mathsf{a}^\dagger = \frac{1}{\sqrt{2}}\left(\frac{\mathsf{Q}}{\lambda} - \frac{i}{\hbar}\mathsf{P}\lambda\right),\quad \mathsf{I},
 \end{equation}
 where $\lambda$ has the same dimensions as $\mathsf{Q}$, $\mathsf{P}$ has dimensions
 of $\hbar/\lambda$ and $\mathsf{a}, \mathsf{a}^\dagger$ satisfy the commutation relation
 \begin{equation}
   \label{eq:CCR_boson}
   [\mathsf{a}, \mathsf{a}^\dagger] = \mathsf{I}.
 \end{equation}

 Correspondingly, the observable quantities of the system can be
 expressed as operator-valued functions of $\mathsf{a}$ and
 $\mathsf{a}^\dagger$. For example, the operator
 $\mathsf{N} = \mathsf{a}^\dagger \mathsf{a}$ together with the
 operators $\mathsf{a}$, $\mathsf{a}^\dagger$, $\mathsf{I}$ generates
 the Lie algebra $\mathfrak{h}(4)$. From the commutators
 \begin{equation}
   \label{eq:ladder_operators}
   [\mathsf{a}, \mathsf{N}] = \mathsf{a}, \quad [\mathsf{a}^\dagger, \mathsf{N}] = -\mathsf{a}^\dagger
 \end{equation}
 follows \cite{KlimovChumakov} that $\mathsf{a}$ and
 $\mathsf{a}^\dagger$ are ladder operators with respect to the
 eigenvectors
 $|n\rangle = (\mathsf{a}^{\dagger})^n|0\rangle/\sqrt{n!}$ of
 $\mathsf{N}$:
 \begin{equation}
   \label{eq:ladder_operators_N_eigenstates}
   \mathsf{a} \,|n\rangle = \sqrt{n}\,|n-1\rangle, \quad \mathsf{a}^\dagger |n\rangle = \sqrt{n+1}\,|n+1\rangle,
 \end{equation}
 and we require that $\mathsf{a}\, |0\rangle = 0$ (because the energy of
 a quantum system must be bounded from below). These eigenvectors form a
 complete and orthonormal ($\langle n|m \rangle = \delta_{nm}$) set
 that spans a Hilbert space $\mathcal{H}$, in which one can represent
 the state of a system described by $\mathfrak{h}(4)$, e.g. the
 quantum harmonic oscillator.

 An eigenvector $|\alpha\rangle$ of the operator $\mathsf{a}$ is
 called a coherent state. The corresponding eigenvalue, $\alpha$, is a
 complex number, since this operator is not self-adjoint. In the basis
 $|n\rangle$, a coherent state is expressed as \cite{KlimovChumakov}
 \begin{equation}
   \label{eq:coherent_state_Fock_basis}
   |\alpha\rangle = e^{-\frac12 |\alpha|^2} \sum_{n=0}^\infty \frac{\alpha^n}{\sqrt{n!}} |n\rangle.
 \end{equation}
 These states play a fundamental role in the study of the
 quantum-classical transition \cite{Zurek_et_al}. This becomes
 apparent with the use of phase space methods, which will be discussed
 in the following section.

\section{Quantum mechanics in phase space}
\label{sec:qm_in_phase_space}

Quantum mechanics can also be formulated in terms of unitary
operators, i.e. operators that satisfy $\mathsf{O}\mathsf{O}^\dagger = \mathsf{O}^\dagger \mathsf{O} = \mathsf{I}$. In this
framework, the operators associated with $\mathsf{Q}$ and $\mathsf{P}$ are
\cite{Tarasov}
\begin{equation}
  \label{eq:Weyl_operators}
  \mathsf{U}(a) = e^{\frac{i}{\hbar}a\mathsf{Q}}, \quad  \mathsf{V}(b) = e^{\frac{i}{\hbar}b\mathsf{P}}.
\end{equation}
Since the argument of the exponential function must be dimensionless,
$a$ has the dimensions of $\mathsf{P}$ and $b$ has the dimensions of
$\mathsf{Q}$. The above operators are a special case of the Weyl
operators (or displacement operators)
\begin{equation}
  \label{eq:displacement_operator}
  \mathsf{D}(a,b) = e^{\frac{i}{\hbar}(a\,\mathsf{Q} \,+\, b\,\mathsf{P})},
\end{equation}
which satisfy the canonical commutation relations in Weyl (or integral) form \cite{Tarasov}
\begin{equation}
  \label{eq:CCR_Weyl}
\mathsf{D}(a_1, b_1)\,\mathsf{D}(a_2, b_2) = e^{-\frac{i}{2\hbar}[a_1b_2 - a_2b_1]}\mathsf{D}(a_1 + a_2, b_1 + b_2).
\end{equation}
In terms of the operators $\mathsf{a}$ and $\mathsf{a}^\dagger$, the displacement
operator is given by \cite{KlimovChumakov}
\begin{equation}
  \label{eq:Displacement_operator_a_a_dagger}
  \mathsf{D}(\alpha) = e^{\,\alpha\, \mathsf{a}^\dagger -\, \alpha^\ast \mathsf{a}} = e^{-\frac12|\alpha|^2} e^{\alpha\, \mathsf{a}^\dagger} e^{-\alpha^\ast \mathsf{a}}.
\end{equation}
The coherent states can also be defined as displaced ground states
$|\alpha\rangle := \mathsf{D}(\alpha) |0\rangle$, as can be seen by
comparing this expression with (\ref{eq:coherent_state_Fock_basis}).

The operators acting on a representation space (such as the Hilbert
space spanned by the vectors $|n\rangle$) belong to a Hilbert space
called Liouville space, in which the scalar product is given by
$\langle \mathsf{A}|\mathsf{B} \rangle =
\mathrm{Tr}(\mathsf{A}^\dagger \mathsf{B})$. The set of displacement
operators $\mathsf{D}(\alpha)$ forms a delta-orthogonal basis of this
space:
$\langle \mathsf{D}(\alpha)|\mathsf{D}(\beta) \rangle = \pi
\delta^{(2)}(\alpha - \beta)$.

In this basis, the density operator (or state operator) $\rho$ of a
quantum system is described by the Weyl characteristic function (also
called ambiguity function) \cite{BarnettRadmore}
\begin{equation}
  \label{eq:def_characteristic_function}
  \chi(\eta, \eta^\ast) = \mathrm{Tr}[\rho\, \mathsf{D}(\eta)],
\end{equation}
which will figure prominently in section 5. The following Fourier
transform of $\chi(\eta, \eta^\ast)$ yields the Wigner function \cite{BarnettRadmore}:
\begin{equation}
  \label{eq:Wigner_function}
W(\xi, \xi^\ast) = \frac{1}{\pi^2} \int \mathrm{d}^2\eta\, \chi(\eta, \eta^\ast) \,e^{\,\xi\eta^\ast -\, \xi^\ast\eta}, \quad \mathrm{d}^2\eta = \mathrm{d}\mathrm{Re}(\eta)\,\mathrm{d}\mathrm{Im}(\eta),
\end{equation}
which can also be defined as
$W(\xi, \xi^\ast) = \mathrm{Tr}[\rho\, \mathsf{\Pi}(\xi)]$, where
\begin{equation}
  \label{eq:wigner_operator}
  \mathsf{\Pi(\xi)} = \frac{2}{\pi}\exp{\bigl[ i\pi (\mathsf{a}^{\dagger} - \xi^\ast)(\mathsf{a} - \xi) \bigr]}
\end{equation}
is the Wigner operator (or displaced parity operator)
\cite{BishopVourdas}, which is self-adjoint and therefore is an
observable, unlike the displacement operator. This implies that the
Wigner function is real, whereas the characteristic function is
complex. The geometrical interpretation of these functions is
discussed in \cite{OzorioAlmeida}. For the experimental determination
of the Wigner function we refer the reader to \cite{HarocheRaimond}.

We remark that it is customary to denote the operators $\mathsf{D}$
and $\mathsf{\Pi}$ with a single complex argument even though $\chi$
and $W$ are functions of two complex variables. This mapping of
operators to functions, also known as a mapping from $q$-numbers to
$c$-numbers, is very useful in the study of the quantum-classical
transition, as will be shown in section 5. For a thorough discussion
of mappings of this kind we refer the reader to
\cite{KlauderSudarshan,AgarwalWolf}.

For a coherent state, $W$ is Gaussian, which can easily be seen using
(\ref{eq:Displacement_operator_a_a_dagger})\textendash(\ref{eq:Wigner_function})
together with the fact that a coherent state is an eigenvector of
$\mathsf{a}$. However, for a so-called cat state
\begin{equation}
  \label{eq:cat_state}
  |\psi\rangle = \mathcal{N}^{1/2}(|\alpha\rangle + |-\alpha\rangle), \quad \mathcal{N}^{-1} = 2(1 + e^{-2|\alpha|^2}),
\end{equation}
$W$ takes negative values:
\begin{equation}
  \label{eq:Wigner_function_cat_state}
  W(\xi, \xi^\ast) = \frac{2\, \mathcal{N}}{\pi} \Bigl[ e^{-2|\xi - \alpha|^2} +\, e^{-2|\xi + \alpha|^2} +\,
  2\, e^{-2|\xi|^2} \cos{(4\, \mathrm{Im}(\alpha^\ast \xi))} \Bigr].
\end{equation}
This is a signature of a non-classical state
\cite{KimNoz}. Moreover, it can be proven that the only
positive-definite Wigner functions describing pure states are
Gaussian \cite{SotoClaverie}. This is one reason why coherent states
are considered the most classical quantum states.

It can be shown \cite{KimNoz} that for a Hamiltonian with a general
quadratic potential
$\mathsf{V}(\mathsf{q}) = a\,\mathsf{q}^2 + b\,\mathsf{q} + c$, the
evolution of the Wigner function is given by the classical Liouville
equation for a probability distribution in phase space:
\begin{equation}
  \label{eq:Liouville_equation}
  \frac{\partial}{\partial t} W(q,p,t) = \frac{\partial H}{\partial q} \frac{\partial W}{\partial p} - \frac{\partial H}{\partial p} \frac{\partial W}{\partial q}.
\end{equation}
However, this does not represent classical behavior unless the Wigner
function is positive. For this reason, $W$ is called a
quasi-probability phase space distribution. Moreover, as a consequence
of Heisenberg's uncertainty principle, the Wigner function cannot have
a width smaller than the size of a Planck cell:
$\Delta q \Delta p \geq \hbar$ \cite{KimNoz}. Therefore, quantum
mechanics introduces a coarse-graining in phase space.

One reason why quasi-probability distributions are useful is that they
enable calculating quantum-mechanical expectation values similarly to
averages in classical statistical mechanics. In particular,
distributions belonging to the Cohen class \cite{Cohen}, which
includes the Wigner function, have the property that integrating them
with respect to one canonical variable yields the probability
distribution of the canonically conjugate variable.

In the literature, it is common to describe the transition to
classical behavior as ``taking the limit $\hbar \rightarrow
0$''. However, this characterization is misleading, since $\hbar$ is a
constant of nature and it cannot be made arbitrarily small. What is
meant by this statement is that one may form a dimensionless parameter
involving $\hbar$ and other physical quantities, such that when this
parameter is made arbitrarily small, a quantum equation reduces to a
classical one (e.g. the Liouville equation above). This shows again
the importance of dimensional considerations in quantum mechanics. We
remark that this limiting procedure does not lead to the vanishing of
negative regions in the Wigner function.

\section{Quantum measurements and the measurement master equation}
\label{sec:phase_space_measurements}

\subsection{Measurement in quantum mechanics}
\label{sec:measurement_quantum_mechanics}

In the axiomatization of quantum mechanics carried out by von Neumann
\cite{vonNeumann}, given an observable with a discrete spectrum
$\mathsf{O} = \sum_n \lambda_n | \lambda_n \rangle \langle \lambda_n
|$, the probability that a measurement of $\mathsf{O}$ yields the
result $\lambda_n$ is
\begin{equation}
  \label{eq:probability_von_Neumann_measurement}
  \mathrm{Prob} (\lambda_n) = \mathrm{Tr} [\rho\, | \lambda_n \rangle
  \langle \lambda_n |],  
\end{equation}
and the state of the system after the measurement is
$\rho = | \lambda_n \rangle \langle \lambda_n |$. If the measurement
result is not known, the system is described by the mixed state
\begin{equation}
  \label{eq:state_after_von_Neumann_measurement}
  \rho = \sum_n \mathrm{Prob} (\lambda_n) | \lambda_n \rangle \langle
  \lambda_n |.
\end{equation}

Instead of associating a projector with each measurement result $n$,
in general one may associate a positive operator $\pi_n$ with
it. These operators form a positive-operator-valued measure (POVM)
\cite{Busch_et_al} and must be such that $\sum_n \pi_n =
\mathsf{I}$. Moreover, each operator $\pi_n$ may be decomposed in
terms of pairs ($\mathsf{A}_k, \mathsf{A}_k^{\dagger}$) of operators:
\begin{equation}
  \label{eq:Kraus_decomposition_effect}
  \pi_n = \sum_k \mathsf{A}_{n k}^{\dagger} \mathsf{A}_{n k}.
\end{equation}
In this framework, the probability that a measurement yields the
result $n$ is $\mathrm{Prob} (n) = \mathrm{Tr} [\rho\, \pi_n]$ and the
state of the system after the measurement is
$\rho = (\mathrm{Tr} [\rho\, \pi_n])^{-1} \sum_k \mathsf{A}_{n k}\,
\rho\, \mathsf{A}_{n k}^{\dagger}$. If the measurement result is not
known, the state of a system after performing a generalized
measurement is given by
\begin{equation}
  \label{eq:mixed_state_generalized_measurement}
  \rho = \sum_n \mathrm{Prob} (n) \frac{\sum_k \mathsf{A}_{n k}\, \rho\, \mathsf{A}_{n
      k}^{\dagger}}{\mathrm{Tr} [\rho\, \pi_n]} = \sum_{n,k} \mathsf{A}_{n k}\, \rho\,
  \mathsf{A}_{n k}^{\dagger}.
\end{equation}
We remark that this formalism is quite general and when the
measurement result can take any real or complex value the sums are
replaced by corresponding integrals.

\subsection{Measurement master equation}
\label{sec:measurement_master_equation}

The dynamics of the state operator of an open quantum system is
described under certain approximations by the
Lindblad--Gorini--Kossakowski--Sudarshan master equation:
\begin{equation}
  \label{eq:Lindblad_equation}
  \frac{\partial \rho}{\partial t} = -\frac{i}{\hbar}[\mathsf{H},\rho] + \frac12\sum_{j=1}^d\kappa_j\Bigl( 2\,\mathsf{R}_j\rho \mathsf{R}_j^\dagger - \mathsf{R}_j^\dagger \mathsf{R}_j\rho - \rho\, \mathsf{R}_j^\dagger \mathsf{R}_j \Bigr),
\end{equation}
where $\mathsf{H}$ is a self-adjoint operator with dimension of
energy, $\mathsf{R}_j$ are arbitrary dimensionless operators and $\kappa_j$ are
non-negative real numbers with dimension of frequency. Here we are
only interested in using this equation and refer the reader to
\cite{BreuerPetruccione} for an in depth discussion of its derivation
and the physical considerations behind it.

The change in the state of a system subject to a generalized
measurement can be modeled as a Poisson process with rate $\gamma$
\cite{Hanson}, as follows. We assume that in a short time interval
$\Delta t$ the probability that a measurement occurs is
$\gamma \Delta t$. If a measurement occurs, then the state at the time
$t + \Delta t$ will be given by
(\ref{eq:mixed_state_generalized_measurement}). Otherwise, the state
at this time results from the unitary evolution of the system, given by
the first term in the right-hand side of
(\ref{eq:Lindblad_equation}). To first order in $\Delta t$, the state
of a system subject to this stochastic process is described at time
$t + \Delta t$ by
\begin{equation}
  \label{eq:Poisson_process_measurement}
  \rho (t + \Delta t) = (1 - \gamma \Delta t) \rho (t) -
  \frac{i}{\hbar} [\mathsf{H}, \rho (t)] \Delta t + \gamma \Delta t \sum_{n, k}
  \mathsf{A}_{n k}\, \rho (t)\, \mathsf{A}_{n k}^{\dagger}.
\end{equation}
In the limit $\Delta t \rightarrow 0$ one obtains the measurement master equation \cite{Cresser_et_al}
\begin{equation}
  \label{eq:measurement_master_equation}
  \frac{\partial \rho}{\partial t} = - \frac{i}{\hbar} [\mathsf{H}, \rho] + \gamma
  \Biggl[ \sum_{n, k} \mathsf{A}_{n k}\, \rho\, \mathsf{A}_{n k}^{\dagger} - \rho \Biggr],
\end{equation}
which can be shown to be of the type (\ref{eq:Lindblad_equation}).

\section{Classicalization of systems with a quadratic Hamiltonian}
\label{sec:classicalization_quadratic_Hamiltonian}

A general quadratic Hamiltonian in the basis $\mathsf{Q}$, $\mathsf{P}$ is of the form:
\begin{equation}
  \label{eq:quadratic_Hamiltonian_Q_P}
  \mathsf{H} = c_1\mathsf{Q}^2 + c_2\mathsf{P}^2 + c_3(\mathsf{Q}\mathsf{P} + \mathsf{P}\mathsf{Q}) + c_4\mathsf{Q} + c_5\mathsf{P}, \quad c_i \in \mathbb{R}.
\end{equation}
The corresponding expression in a dimensionless basis analogous to
(\ref{eq:CCR_ladder_basis}) is:
\begin{equation}
  \label{eq:quadratic_Hamiltonian_a_a_dagger}
  \mathsf{H} = z_1\mathsf{b}^\dagger \mathsf{b} + z_2\mathsf{b}^2 + z_2^\ast (\mathsf{b}^\dagger)^2 + z_3\mathsf{b} + z_3^\ast \mathsf{b}^\dagger, \quad z_1 \in \mathbb{R},\quad z_2,z_3 \in \mathbb{C}.
\end{equation}
Using a canonical transformation (see \ref{sec:diagonalization_quadratic_Hamiltonian}) one obtains from
(\ref{eq:quadratic_Hamiltonian_a_a_dagger}) the harmonic
oscillator-like Hamiltonian ($z_0 = (z_1^2 - 4|z_2|^2)^{1/2}$, $\phi = \mathrm{Arg}(z_2)$):
\begin{equation}
  \label{eq:harmonic_oscillator_hamiltonian}
  \mathsf{H} = z_0\, \mathsf{a}^\dagger \mathsf{a} + c\,\mathsf{I},
\end{equation}
with
\begin{equation*}
  \label{eq:ladder_operator_diagonal_basis}
    \mathsf{b} = \biggl[\frac12 \frac{z_1 + z_0}{z_0} \biggr]^{1/2} e^{i\phi/2}\, \mathsf{a} + \biggl[\frac12 \frac{z_1 - z_0}{z_0} \biggr]^{1/2} e^{i\phi/2} \, \mathsf{a}^\dagger + \frac{2 z_2 z_3^\ast - z_1 z_3}{z_0^2}\mathsf{I}
  \end{equation*}
  and
  \begin{equation*}
    \label{eq:ground_state_energy_diagonalized_Hamiltonian}
    c = \frac12(z_0 - z_1) + \frac{1}{z_0^2}\left(z_2 (z_3^\ast)^2 + z_2^\ast z_3^2 - z_1 |z_3|^2\right).
  \end{equation*}
  We note that the constant $c$ in the Hamiltonian does not affect the
  dynamics and, therefore, in this basis the system behaves like a
  harmonic oscillator with ground-state energy zero. This is a
  demonstration of the usefulness of algebraic methods in quantum
  mechanics. In the following we will set $z_0 = \hbar\,\omega$.

\subsection{Simultaneous imprecise measurement of the quadratures of the harmonic oscillator}
\label{measurement_of_quadratures}

For the quantum harmonic oscillator
(\ref{eq:harmonic_oscillator_hamiltonian}) one can define the
dimensionless self-adjoint quadrature operators:
\begin{equation}
  \label{eq:quadratures}
\mathsf{X} = \frac12 (\mathsf{a} + \mathsf{a}^{\dagger}), \quad \mathsf{Y} = \frac{1}{2i}(\mathsf{a} - \mathsf{a}^{\dagger}), \quad  [\mathsf{X}, \mathsf{Y}] = \frac{i}{2}.
\end{equation}
For any state of the harmonic oscillator, the Heisenberg uncertainty
product is given by $\Delta \mathsf{X} \,\Delta \mathsf{Y} \geq 1 / 4$,
with
$\Delta \mathsf{O} = (\langle \mathsf{O}^2 \rangle - \langle
\mathsf{O} \rangle^2)^{1/2}$. States that satisfy the equality are
called minimum-uncertainty states. For example, coherent states have
this property: $ \Delta \mathsf{X} = \Delta \mathsf{Y} =
\frac{1}{2}$. This is another reason why coherent states are
considered the most classical quantum states. The eigenvalues of
$\mathsf{X}$ and $\mathsf{Y}$ are the real and imaginary parts of the
amplitude $\alpha = \alpha_x + i \alpha_y$. A measurement of $\alpha$
can be described using the formalism of generalized measurements of
section 4, as follows.

Given a positive operator with unit trace $\Psi$, one can define the
operators \cite{Walker}:
\begin{equation}
  \label{eq:POVM_quadrature_measurement}
  \pi_{\alpha} = \frac{1}{\pi} \mathsf{D} (\alpha) \Psi \mathsf{D}^{\dagger} (\alpha), \quad
  \int \mathrm{d}^2\alpha\, \pi_{\alpha} = \mathsf{I}.
\end{equation}
The probability density of obtaining the result $\alpha$ is
$\mathrm{Prob}(\alpha) = \mathrm{Tr} [\pi_{\alpha}\,\rho]$, and its moments are given by
\begin{equation}
  \label{eq:moments_distribution_alpha}
  M_{i j} = \int \mathrm{d}^2\alpha\, \mathrm{Prob}(\alpha)\, \alpha_x^i \,\alpha_y^j , \quad \mathrm{d}^2\alpha = \mathrm{d} \alpha_x \, \mathrm{d} \alpha_y.
\end{equation}
The expectation values of $\alpha_x$ and $\alpha_y$ are given by
\begin{eqnarray}
  \label{eq:expectation_values_quadratures}
  \langle \alpha_x \rangle & = & M_{10} = \langle \mathsf{X} \rangle_\rho - \langle \mathsf{X} \rangle_\Psi,\nonumber\\
  \langle \alpha_y \rangle & = & M_{01} = \langle \mathsf{Y} \rangle_\rho - \langle \mathsf{Y} \rangle_\Psi,
\end{eqnarray}
so in order for the measurement to be unbiased, the expectation values
with respect to $\Psi$ must be zero. The second-order moments give the
mean square uncertainty for the simultaneous measurement of $\alpha_x$
and $\alpha_y$:
\begin{eqnarray}
  \label{eq:second_moments_quadratures}
  \langle (\Delta \alpha_x)^2 \rangle & = & M_{20} - (M_{10}^{})^2 = \langle
  \mathsf{X}^2 \rangle_{\rho} + \langle \mathsf{X}^2 \rangle_{\Psi},\nonumber\\
  \langle (\Delta \alpha_y)^2 \rangle & = & M_{02} - (M_{01}^{})^2 = \langle
  \mathsf{Y}^2 \rangle_{\rho} + \langle \mathsf{Y}^2 \rangle_{\Psi},
\end{eqnarray}
which shows that the measurement contributes (classical noise) to the
uncertainty (quantum noise) of the measured state. Therefore, it is an
imprecise, simultaneous measurement of two non-commuting
observables. In the following we will show that this interaction
between the harmonic oscillator and the system described by $\Psi$
leads to classicalization of the former.

\subsection{Measurement-induced classicalization}
\label{sec:measurement_induced_classicalization}

From the theory of generalized measurements in section
\ref{sec:phase_space_measurements}, the operators
(\ref{eq:POVM_quadrature_measurement}) can be written as:
\begin{equation}
  \label{eq:Kraus_decomposition_pi}
  \pi_\alpha = \mathsf{A}^\dagger_\alpha \mathsf{A}_\alpha,\quad \mathsf{A}_\alpha = \frac{1}{\sqrt{\pi}} \, \mathsf{D}(\alpha) \Psi^{1/2} \mathsf{D}^\dagger(\alpha).
\end{equation}
In turn, $\mathsf{A}_\alpha$ can be written in the basis of displacement operators:
\begin{equation}
  \label{eq:effect_basis_displacement_ops}
\fl \mathsf{A}_\alpha = \int \mathrm{d}^2\beta\, \mathsf{D}(\beta) \mathrm{Tr}[\mathsf{D}^\dagger(\beta) \mathsf{A}_\alpha]
          = \frac{1}{\sqrt{\pi}}\int \mathrm{d}^2\beta\, \mathsf{D}(\beta)\, e^{\,\beta^\ast \alpha \,-\, \alpha^\ast \beta} \chi^\ast_{\Psi^{1/2}}(\beta),
\end{equation}
where we used the cyclic property of the trace, the definition
(\ref{eq:def_characteristic_function}) and the identity
\begin{equation}
  \label{eq:displacement_of_displacement_operator}
  \mathsf{D}^\dagger(\xi) \mathsf{D}(\zeta) \mathsf{D}(\xi) = \mathsf{D}(\zeta) \exp{(\xi^\ast \zeta - \zeta^\ast \xi)}.
\end{equation}
In this representation,
\begin{equation}
  \label{eq:twirling_superoperator}
  \int \mathrm{d}^2\alpha\, \mathsf{A}_\alpha \,\rho\, \mathsf{A}_\alpha^\dagger =
  \pi \int \mathrm{d}^2\gamma\, \mathsf{D}(\gamma)\, \rho\, \mathsf{D}^\dagger(\gamma)\, |\chi_{\Psi^{1/2}}(\gamma)|^2,    
\end{equation}
where we used the identity
\begin{equation}
  \label{eq:2d_Dirac_delta}
  \delta^{(2)}(\beta\, -\, \gamma) = \frac{1}{\pi^2} \int_{-\infty}^{\infty} \mathrm{d} ^2\alpha \exp[\alpha\beta^* - \alpha^*\beta] \exp[\alpha^*\gamma - \alpha\gamma^*].
\end{equation}
Defining the even, positive-definite function
$g(|\alpha|) := \mathcal{N}\,|\chi_{\Psi^{1/2}}(\alpha)|^2$, where
$\mathcal{N}$ is a normalization constant, we finally arrive at the
master equation corresponding to the measurement described above:
\begin{equation}
  \label{eq:master_equation_quadrature_measurement}
\frac{\partial \rho}{\partial t} = - \frac{i}{\hbar} [\mathsf{H}, \rho] + \gamma
   \int \mathrm{d}^2\alpha\, g (| \alpha |) [\mathsf{D} (\alpha) \,\rho\, \mathsf{D}^{\dagger} (\alpha) -
   \rho],  
\end{equation}
where $\mathsf{H}$ is defined in
(\ref{eq:harmonic_oscillator_hamiltonian}).  The integral term can be
interpreted as describing phase space ``kicks'' occurring with rate
$\gamma$ and with a $g$-distributed strength $\alpha$.

Taking the trace of (\ref{eq:master_equation_quadrature_measurement})
multiplied with $\mathsf{D}(\eta)$ yields the following mapping from
operators to functions:
\begin{eqnarray}
  \label{eq:Bargmann_rep_superoperators}
  \mathsf{a}^\dagger \mathsf{a} \,\rho \rightarrow \biggl( - \frac{\partial}{\partial \eta} \frac{\partial}{\partial \eta^\ast} + \eta^\ast \frac{\partial}{\partial \eta^\ast} \biggr) \chi_\rho(\eta, \eta^\ast, t),\nonumber\\
  \rho \, \mathsf{a}^\dagger \mathsf{a} \rightarrow \biggl( - \frac{\partial}{\partial \eta} \frac{\partial}{\partial \eta^\ast} + \eta \frac{\partial}{\partial \eta} \biggr) \chi_\rho(\eta, \eta^\ast, t),\nonumber\\
  \int \mathrm{d}^2\alpha\, g(|\alpha|)\,\mathsf{D}(\alpha) \,\rho\, \mathsf{D}^\dagger(\alpha) \rightarrow \chi_\rho(\eta, \eta^\ast, t) \,\chi_g(|\eta|),
\end{eqnarray}
and the corresponding dynamic equation for the characteristic function:
\begin{equation}
  \label{eq:equation_char_fun_stationary_frame}
  \biggl( \frac{\partial}{\partial t} + i\omega[\eta^\ast \partial_{\eta^\ast} - \eta \partial_\eta] + \gamma[1 - \chi_g(|\eta|)]  \biggr)\chi_\rho(\eta, \eta^\ast, t) = 0.
\end{equation}
Introducing the function
$\tilde{\chi}_\rho(\eta, \eta^\ast, t) = \chi_\rho(\eta e^{i \omega t},
\eta^\ast e^{-i \omega t} t)$, the equality of the total
differentials yields:
\begin{equation}
  \label{eq:time_derivative_tilde}
  \frac{\partial \tilde{\chi}_\rho}{\partial t} = i\omega\eta \frac{\partial \chi_\rho}{\partial \eta} - i\omega\eta^\ast \frac{\partial \chi_\rho}{\partial \eta^\ast} + \frac{\partial \chi_\rho}{\partial t},
\end{equation}
from which one obtains the equation of motion for the characteristic
function of the state in a frame rotating with frequency $\omega$:
\begin{equation}
  \label{eq:equation_characteristic_function}
\frac{\partial}{\partial t} \tilde{\chi}_{\rho} (\eta, \eta^\ast, t) = -\, \gamma [1 - \chi_g (| \eta |)] \,\tilde{\chi}_{\rho} (\eta,\eta^\ast, t),
\end{equation}
which in the non-rotating frame has the solution:
 \begin{equation}
   \label{eq:solution_characteristic_function}
   \chi_{\rho} (\eta, \eta^\ast, t) = \chi_0 (\eta e^{i \omega t}, \eta^\ast e^{-i \omega t}) \exp \{ -\, \gamma t[1
   - \chi_g (| \eta |)] \}.
 \end{equation}
 For economy of notation, in the following we omit the second argument
 of $\chi_\rho$. We recognize the exponential term as the
 characteristic function of a compound Poisson process with rate
 $\gamma t$ and jump-size distribution $g$ \cite{Hanson}. Since
 $\chi_g$ is the characteristic function of a probability
 distribution, it satisfies $\chi_g(0)=1$ and in the limit
 $|\eta| \rightarrow \infty$ it vanishes for all $t$:
 $\chi_g(|\eta|) \rightarrow 0$. Therefore, at any time the
 exponential term is an even function that asymptotically decays in
 phase space towards the plane $f(\eta) = e^{-\gamma t}$, which in
 turn decays with time towards zero.
 
 \begin{figure}[t]
   \begin{tabular}{@{}cc}
     \includegraphics[scale=1.2]{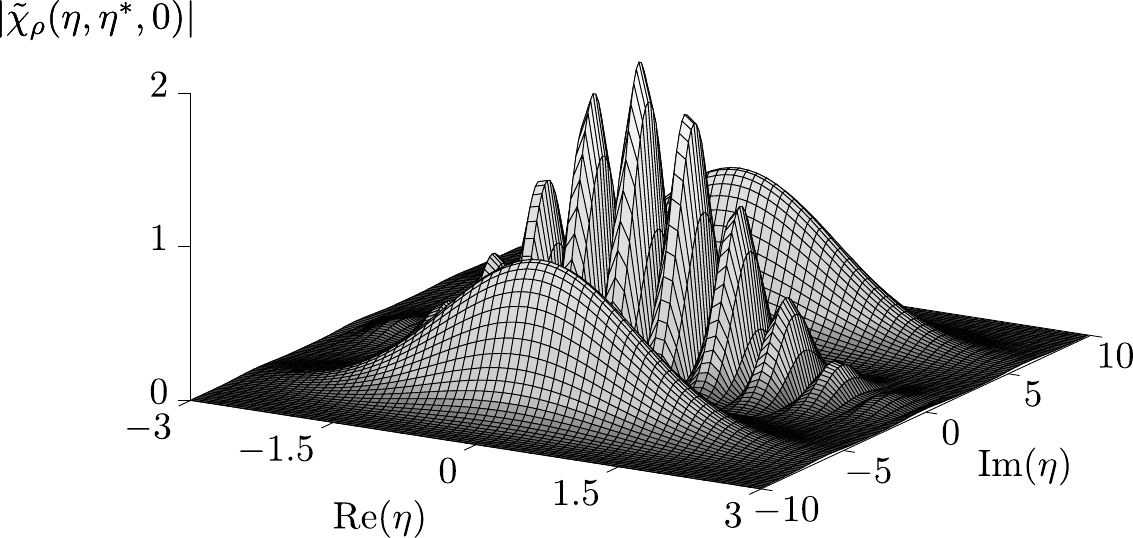}\\
     \includegraphics[scale=1.2]{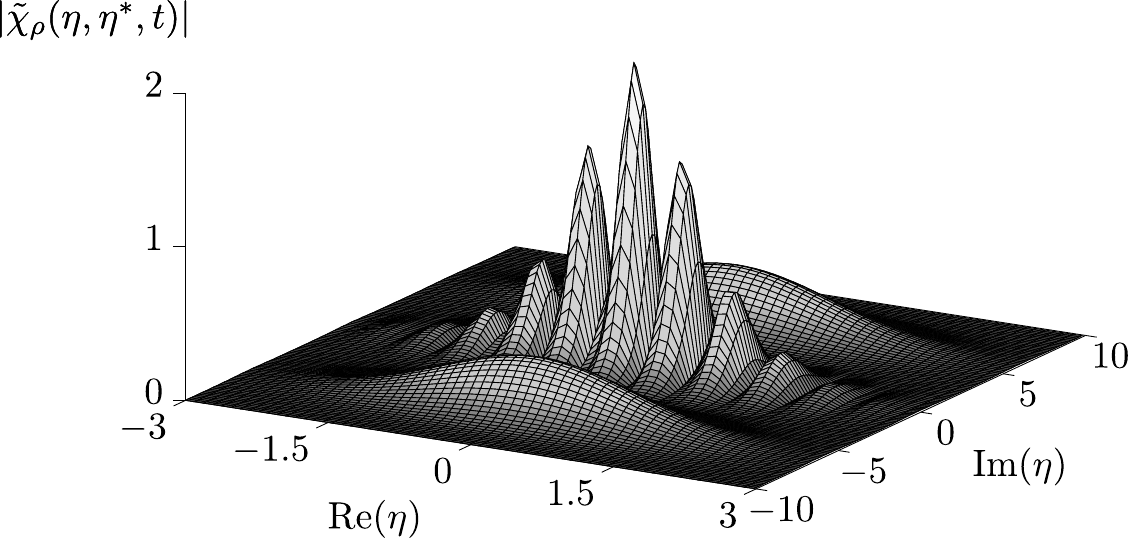}
     \end{tabular}
     \caption{Top: $|\chi(\eta,0)|$ for the cat state
       (\ref{eq:Wigner_function_cat_state}) with $\alpha=3i$. Bottom:
       Snapshot of $|\chi(\eta,t)|$ at $\gamma t = 1$ for the cat
       state subject to an imperfect, simultaneous measurement of the
       quadratures assuming $g$ to be a Gaussian distribution. The
       strong damping of the outer peaks indicates that the
       measurement is classicalizing the state.}
   \label{fig:char_fun_cat_state_w_noise_wo_noise}
 \end{figure}
 
 Let us assume that $\chi_0$ corresponds to the cat state
 (\ref{eq:Wigner_function_cat_state}). Since $\chi_\rho$ is a complex
 function related to the Wigner function through a Fourier transform,
 we can interpret $|\chi_\rho(\eta, t)|$ as a phase space Fourier
 spectrum \cite{Chountasis_et_al}. For $t=0$, it is shown in the top
 panel in figure \ref{fig:char_fun_cat_state_w_noise_wo_noise}. The
 central, structured peak corresponds to the low frequencies of the
 Wigner function, whereas the two outer peaks correspond to the high
 frequencies, which are associated to the non-classical character of
 the state.

 A snapshot of the dynamics given by
 (\ref{eq:equation_characteristic_function}) is shown in the bottom
 panel for $\gamma t = 1$. For illustration purposes, the distribution
 $g$ is assumed to be Gaussian. The noticeable suppression of the
 outer peaks is due to the ``filtering'' effect of the measurement,
 described by the decaying exponential term in
 (\ref{eq:solution_characteristic_function}). Moreover, repeating the
 above analysis for a coherent state reveals that it will be
 broadened, as expected from (\ref{eq:second_moments_quadratures}).
 Therefore, it is clear that the measurement drives a superposition
 state of the harmonic oscillator towards a mixture of Gaussian
 states, which corresponds to a classical phase space function.

 It is interesting to consider the case of very frequent and very
 small phase-space kicks, which amounts to assuming that
 $\gamma \rightarrow \infty$ and that the second moments of the
 distribution $g$ are much larger than the higher moments, but their
 product with $\gamma$ remains finite. Performing a series expansion
 of the displacement operator to second order in $\eta$:
\begin{equation}
  \label{eq:series_displacement_op_second_order}
  \mathsf{D}(\eta) \approx \mathsf{I} + \eta \mathsf{a}^\dagger - \eta^\ast \mathsf{a} + \frac12 (\eta \mathsf{a}^\dagger - \eta^\ast \mathsf{a})(\eta \mathsf{a}^\dagger - \eta^\ast \mathsf{a}),
\end{equation}
one obtains from (\ref{eq:master_equation_quadrature_measurement}) a
master equation describing diffusion in phase space \cite{Agarwal}:
\begin{equation}
  \label{eq:master_equation_diffusion}
\fl \frac{\partial \rho}{\partial t} = -i\omega [\mathsf{a}^\dagger \mathsf{a},\rho] - \kappa(-2\,\mathsf{a}\, \rho\, \mathsf{a}^\dagger + \mathsf{a}^\dagger \mathsf{a} \,\rho + \rho\, \mathsf{a}^\dagger \mathsf{a}  - 2\,\mathsf{a}^\dagger \rho\, \mathsf{a} + \mathsf{a}\, \mathsf{a}^\dagger \rho + \rho\, \mathsf{a}\, \mathsf{a}^\dagger).
\end{equation}
In fact, one can describe this dynamics in terms of the Brownian
motion of the state vector of the system in Hilbert space
\cite{JacobsSteck}. This and other master equations describing
Gaussian dynamics are thoroughly discussed in \cite{Serafini}. It is
interesting to note that under this kind of evolution, the Wigner
function becomes positive everywhere in a finite time
\cite{BrodierOzorio}, in contrast to the example discussed here.

\section{Conclusions}
\label{sec:conclusions}

The model presented in section 5 illustrates the measurement approach
to the classicalization of a quantum system by means of a master
equation that can be solved exactly for an arbitrary measurement
strength (``kick'' size). Throughout the article, we aimed at keeping
the presentation general and, at the same time, accessible. This
should enable newcomers to the field to apply the measurement approach
to classicalization to more complex systems.

\ack

The author acknowledges the kind support of a CONACYT-DAAD PhD
scholarship and is grateful to Klaus Hornberger for his suggestions to
improve the quality of the manuscript.

\appendix

\section{Diagonalization of a quadratic Hamiltonian}
\label{sec:diagonalization_quadratic_Hamiltonian}

Following \cite{Zelevinsky}, we consider the canonical transformation
\begin{equation}
  \label{eq:bogoliubov_trafo}
  \mathsf{b} = \mu \mathsf{a} + \nu \mathsf{a}^\dagger + \delta.
\end{equation}
From
$[\mathsf{a}, \mathsf{a}^\dagger] = [\mathsf{b}, \mathsf{b}^\dagger] =
\mathsf{I}$, follows that $|\mu|^2 - |\nu|^2 = 1$. Substituting in
(\ref{eq:quadratic_Hamiltonian_a_a_dagger}) and collecting terms, in
order to get a diagonal operator the following equations must be
satisfied:
\begin{eqnarray}
  \label{eq:equations_diagonalization}
  z_1 \nu^\ast \delta + z_1 \delta^\ast \mu + 2 z_2^\ast \mu \delta + 2 z_2 \nu^\ast \delta^\ast + z_3^\ast \mu + z_3 \nu^\ast = 0\\
  z_1 \mu^\ast \delta + z_1 \delta^\ast \nu + 2 z_2^\ast \nu \delta + 2 z_2 \mu^\ast \delta^\ast + z_3^\ast \nu + z_3 \mu^\ast = 0\\
  z_1 \nu^\ast \mu + z_2^\ast \mu^2 + z_2 (\nu^\ast)^2 = 0.
\end{eqnarray}
Subtracting (A.2) multiplied by $\nu$ from (A.3) multiplied by $\mu$,
yields $\delta = -(2 z_2 \delta^\ast + z_3)/z_1$. Substituting in
(A.2) we obtain $\delta^\ast = (2 z_2^\ast z_3 - z_1 z_3^\ast)/z_0^2$,
with $z_0 = (z_1^2 - 4|z_2|^2)^{1/2}$. From this expression we arrive
at the condition $z_1 > 2|z_2|$.

In order to find $\mu$ and $\nu$ from (A.4) we use the polar
representations
\begin{equation}
  \label{eq:polar_reps}
  z_2 = |z_2| e^{i \phi}, \quad \mu = U e^{i \phi_u}, \quad \nu = V e^{i \phi_v},
\end{equation}
and choose $\phi_u = \phi_v = \frac12 \phi$ in order to obtain an
equation with real variables. We now use the parametrizations
$U = \cosh(\Theta/2)$ and $V = \sinh(\Theta/2)$, and recall the
identities
\begin{equation}
  \label{eq:hyperbolic_identities}
  \cosh^2(x) + \sinh^2(x) = \cosh(2x),\quad 2\sinh(x)\cosh(x) = \sinh(2x).
\end{equation}
In terms of $\Theta$, (A.4) has the form
$(z_1 \sinh \Theta)/2 + |z_2| \cosh \Theta = 0$. Using
$2\,\mathrm{atanh}(x) = \log[(1+x)/(1-x)]$, we obtain
$\Theta = \log \sqrt{(z_1 - 2|z_2|)/(z_1 + 2|z_2|)}$. Now we can
calculate $\sinh \Theta = -2|z_2|/z_0$ and $\cosh \Theta =
z_1/z_0$. Recalling the identities
\begin{equation}
  \label{eq:more_identities}
  \cosh \frac{x}{2} = \sqrt{(\cosh x + 1)/2}, \quad   \sinh \frac{x}{2} = \sqrt{(\cosh x - 1)/2},
\end{equation}
we finally arrive at
\begin{equation}
  \label{eq:U_and_V}
  U = \Bigl[ \frac12\, \frac{z_1 + z_0}{z_0} \Bigl]^{1/2}, \quad   V = \Bigl[ \frac12\, \frac{z_1 - z_0}{z_0} \Bigl]^{1/2}.
\end{equation}
Substituting all of the above in
(\ref{eq:quadratic_Hamiltonian_a_a_dagger}), one obtains
(\ref{eq:harmonic_oscillator_hamiltonian}).

\bibliography{classicalization_refs}

\end{document}